# Critical behavior of one-component system with Yukawa interaction potential

*I.K. Loktionov*

*Donetsk National Technical University,
58 Artioma Str., Donetsk 38000, Ukraine*
*likk@telenet.dn.ua*
math@dgtu.donetsk.ua

**Abstract**

It is shown that the liquid-gas phase transition exists in one-component system with potential Yukawa interaction. The value of the critical exponent $\beta$ of the order parameter is found. The relation between the potential parameters of Yukawa interaction and critical values of density, temperature and pressure is obtained.

## 1. Introduction

Recently [1] critical exponent $\beta = 1/3$ in the vicinity of critical point of phase transition in pure one-component systems has been found by semi-phenomenological (generalized lattice) model. The approach used [1] is quite simple and transparent, however it is not rigorous enough. On the other hand, few years ago [2] a method of factorization of the exact configuration integral of the system of $N$ interacting particles which application Weyl theorem has been used to derive the partition function:

$$\ln Z = \frac{V}{2}\left\{\int_{\Omega^+}\frac{d^D k}{(2\pi)^D}\ln\left[\frac{VT}{2v^+(k)}\int_0^{+\infty}\exp\left(-\frac{VT\rho^2}{4v^+(k)}\right)J_0^N(\rho)\rho d\rho\right] + \right.$$
$$\left. + \int_{\Omega^-}\frac{d^D k'}{(2\pi)^D}\ln\left[\frac{VT}{2v^-(k')}\int_0^{+\infty}\exp\left(-\frac{VT\rho^2}{4v^-(k')}\right)I_0^N(\rho)\rho d\rho\right]\right\} + N\left(\frac{v_0}{2T} + \ln V\right) \quad (1)$$

for the system with pair central potential $v(|r|)$ admitting an expansion into Fourier series. Here $D$ is the space dimensionality, $v^\pm(k)$ are the positive and negative Fourier transform.

The idea of the integrand factorization was used in the work [3], where the configuration integral for the systems considered was calculated by the saddle-point approximation. In the quadratic approximation auxiliary variable $\rho$, the contribution of interatomic potentials into the Helmholtz free energy was found

$$F = -T\ln Z = F_{id} - \frac{N}{2}(v_0 - n\widetilde{v}_0) + \frac{VT}{2}\int_\Omega \frac{d^D k}{(2\pi)^D}\ln(1 + n\widetilde{v}(k)/T), \quad (2)$$

where $D$ is the dimensionality of space, $n = N/V$ is the density, $v(r)$ and $\tilde{v}(k)$ are the pair central interatomic potential and its Fourier transform, respectively, $v_0 = v(0)$ is the value of the potential at $r = 0$ and $\tilde{v}_0 = \tilde{v}(0)$ is the value of Fourier transform with $k = 0$. It is important to note that Eq.(2) follows from Eq.(1) if the calculation of internal integrals containing Bessel functions (using Laplace method) is restricted by the in quadratic approximation with respect $\rho$. Furthermore, Eq.(2) coincides with the equation for free energy in [4] where the method of collective variables is used.

The goal of this paper is to refine a study of the saturation curve in a small vicinity of the critical point of liquid-gas phase transition performed in [1] by more rigorous approach, based on exact partition function calculation.

## 2. Thermodynamic potential

Let us considers a three-dimensional model with repulsive Yukawa pair potential of the from

$$v(r) = \frac{A}{4\pi r}\exp(-ar), \quad (A > 0, a > 0). \tag{3}$$

Its Fourier transform has the form

$$\tilde{v}(k) = \frac{A}{k^2 + a^2}. \tag{4}$$

The integration over $k$ in the right hand side Eq.(2) with Fourier transform Eq.(4) leads to the following free energy

$$F = F_{id} + \frac{N^2 w}{2V} + \frac{Na^3 w}{8\pi} + \frac{VTa^3}{12\pi}\left[1 - \left(\sqrt{1 + nw/T}\right)^3\right], \tag{5}$$

where $w = \tilde{v}(0) = A/a^2$. Note, the singularity of the Yukawa potential at $r = 0$ compensated by the divergent part of the integral over $k$ in (2). Using this free energy and standard relations of thermodynamics one can easily find the pressure $P$ and chemical potential $\mu$:

$$P = nT + \frac{n^2 w}{2} - \frac{a^3 T}{12\pi}\left[1 - \sqrt{1 + nw/T}\left(1 - \frac{nw}{2T}\right)\right], \tag{6}$$

$$\mu = T\left(\ln(n \cdot \lambda^3) + \frac{nw}{T} + \frac{a^3 w}{8\pi T}\left(1 - \sqrt{1 + nw/T}\right)\right) \tag{7}$$

The isotherms for this equation of state have characteristic Van der Waals loops. The relation between the critical temperature $T_c$ and density $n_c$ and the potential parameters $A, a$ have the following form

$$n_c = \frac{a^3}{12\pi\sqrt{3}}, \quad T_c = \frac{a^3 w}{24\pi\sqrt{3}}, \quad P_c = (2-\sqrt{3})n_c k_b T_c, \quad n_c \beta_c w = 2.$$

## 3. The behavior of substance in vicinity of the critical point

Under the conditions of thermodynamic phase equilibrium

$$\begin{cases} P(n_1,T) = P(n_2,T), \\ \mu(n_1,T) = \mu(n_2,T), \end{cases} \tag{8}$$

where $n_1$ and $n_2$ are phases densities at the temperature $T < T_c$. Using formulae (6) and (7), let us expand Eq. (8) into Taylor series in powers of the small order parameter $\eta = (n_2 - n_1)/2$ near the critical point. This results in the system of equations analogous to the system obtained in work [1]:

$$\begin{cases} B_1 \eta + B_2 \eta^3 = 0, \\ C_1 \eta + C_2 \eta^3 = q\theta. \end{cases} \tag{9}$$

But, in the contrast to [1] the obtained coefficients of the equation system Eq.(9) depend on the temperature:

$$B_1 = \frac{6w}{T}\frac{(\xi+1)^2 + C\xi\sqrt{\xi+1}}{\xi+1}, \quad C_1 = \frac{(3\xi+2)}{6\xi}B_1, \quad C = -\frac{a^3 w}{16\pi T},$$

$$B_2 = -C\left(\frac{w}{T}\right)^3 \frac{(2\xi+3)\sqrt{\xi+1}}{4(\xi+1)^3}, \quad C_2 = \left(\frac{w}{T}\right)^3 \left[\frac{2}{3\xi^3} - C\frac{(\xi+2)\sqrt{\xi+1}}{8(\xi+1)^3}\right], \quad \xi = \frac{w}{2T}(n_1 + n_2),$$

$\theta = 1 - T/T_c$ is the dimensionless temperature, the value $q = \Delta H_p/N_A$ is determined by the transition heat $\Delta H_p$.

In the critical point $C_1/B_1 = (3\xi+2)/6\xi$ and $B_1 = C_1 = 0$. Let us multiply of the first equation of the Eq.(9) by $-(3\xi+2)/6\xi$ and sum it with the second one. It leads to the relation

$$\eta = K \cdot \theta^{1/3}, \text{ where } K = \left(\frac{6q\xi}{6\xi C_2 - (3\xi+2)B_2}\right)^{1/3}.$$

The values $\xi, B_2, C_2$ here depend on the temperature. However, in extremely narrow vicinity of the critical point the only zero order term of the expansion is important. So, all the further orders with respect to $\theta$ may be neglected in the expansions of the values $\xi, B_2, C_2$. In leads to

$$\eta = K_0 \cdot \theta^{1/3}, \tag{10}$$

where $K_0 = \left(\dfrac{6q\xi_0}{6\xi_0 C_{20} - (3\xi_0 + 2)B_{20}}\right)^{1/3}$, $\xi_0 = \dfrac{w}{2T_c}(n_1 + n_2)$, $C_0 = -\dfrac{a^3 w}{16\pi T_c}$,

$$B_{20} = -C_0 \left(\frac{w}{T_c}\right)^3 \frac{(2\xi_0+3)\sqrt{\xi_0+1}}{4(\xi_0+1)^3}, \quad C_{20} = \left(\frac{w}{T_c}\right)^3 \left[\frac{2}{3\xi_0^3} - C_0 \frac{(\xi_0+2)\sqrt{\xi_0+1}}{8(\xi_0+1)^3}\right].$$

It is seen directly from (10) that the critical exponent $\beta$ of the order parameter in the one-component system with Yukawa potential is equal to $1/3$ that coincides with the parameter value for the model considered in [1].

## 5. Conclusion

Probably, this result is not restricted by the simple Yukawa potential. Analogous results can be obtained for the systems with more complex but realistic potentials as well. In particular, the similar results take a place for the following potentials:

$$v(r) = \frac{\exp(-ar)}{4\pi}\left(\frac{A}{r} - \frac{B}{2a}\right) \text{ и } v(r) = \frac{1}{4\pi r}\left(A\exp(-ar) - B\exp(-ar)\right).$$

These potential have typical potential well between repulsion at small distances and attraction at large ones as well as positive Fourier transform $\tilde{v}(k)$ at the conditions $0 < B/Aa^2 < 1$ and $0 < B/A < (b/a)^2$. Accounting this results, one can believe that the critical exponent found in this approach does not depend on a specific form of the potential.

**Acknowledgements**

The author would like to thank Professor S.V. Terekhov for the fruitful discussion of the results and useful advice while preparing the article.